\documentclass[12pt]{article}
\usepackage{graphicx}
\usepackage{mathrsfs}
\usepackage{color}
\usepackage{cite}
\usepackage{amsmath,slashed, amssymb,amsbsy}
\usepackage{epstopdf}
\usepackage{multicol}
\usepackage{setspace}
\usepackage{fancyhdr}
\usepackage{hyperref}
\usepackage[titletoc]{appendix}
\numberwithin{equation}{section} 

\usepackage{axodraw}

\def\beq{\begin{eqnarray}}
\def\eeq{\end{eqnarray}}
\def\bea{\begin{eqnarray*}}
\def\eea{\end{eqnarray*}}
\newenvironment{Eqnarray}%
     {\arraycolsep 0.14em\begin{eqnarray}}{\end{eqnarray}}
\newcommand{\beqa}{\begin{Eqnarray}}
\newcommand{\eeqa}{\end{Eqnarray}}
\def\eq#1{eq.~(\ref{#1})}
\def\eqs#1#2{eqs.~(\ref{#1}) and (\ref{#2})}
\def\Eqss#1#2#3{Eqs.~(\ref{#1}), (\ref{#2}) and (\ref{#3})}
\def\half{\tfrac{1}{2}}

\def\nn{\nonumber}
\def\sigmabar{\overline{\sigma}}
\def\ahat{\hat{a}}

\def\dslash{\not{\hbox{\kern-2pt $\partial$}}}
\def\Dslash{\not{\hbox{\kern-4pt $D$}}}
\def\Oslash{\not{\hbox{\kern-4pt $O$}}}
\def\Qslash{\not{\hbox{\kern-4pt $Q$}}}
\def\pslash{\not{\hbox{\kern-2.3pt $p$}}}
\def\lpslash{\not{\hbox{\kern-3.2pt $p$}}}
\def\kslash{\not{\hbox{\kern-2.3pt $k$}}}
\def\qslash{\not{\hbox{\kern-2.3pt $q$}}}
\def\deltaxi{\delta_{\eta}}
\def\phm{\phantom{-}}
\def\vev#1{\left\langle #1\right\rangle}

\def\ifmath#1{\relax\ifmmode #1\else $#1$\fi}
\def\dalpha{\dot{\alpha}}
\def\dbeta{\dot{\beta}}
\def\centeron#1#2{{\setbox0=\hbox{#1}\setbox1=\hbox{#2}\ifdim
\wd1>\wd0\kern.5\wd1\kern-.5\wd0\fi
\copy0\kern-.5\wd0\kern-.5\wd1\copy1\ifdim\wd0>\wd1
\kern.5\wd0\kern-.5\wd1\fi}}
\def\ltap{\;\centeron{\raise.35ex\hbox{$<$}}{\lower.65ex\hbox{$\sim$}}\;}
\def\gtap{\;\centeron{\raise.35ex\hbox{$>$}}{\lower.65ex\hbox{$\sim$}}\;}

\def\ls#1{\ifmath{_{\lower1.5pt\hbox{$\scriptstyle #1$}}}}
\def\newcdot{\kern.06em{\cdot}\kern.06em}

\def\slashchar#1{\setbox0=\hbox{$#1$}           % set a box for #1
   \dimen0=\wd0                                 % and get its size
   \setbox1=\hbox{/} \dimen1=\wd1               % get size of /
   \ifdim\dimen0>\dimen1                        % #1 is bigger
      \rlap{\hbox to \dimen0{\hfil/\hfil}}      % so center / in box
      #1                                        % and print #1
   \else                                        % / is bigger
      \rlap{\hbox to \dimen1{\hfil$#1$\hfil}}   % so center #1
      /                                         % and print /
   \fi}                                        %

\def\singleandthirdspaced{\baselineskip=\normalbaselineskip\multiply
    \baselineskip by 130\divide\baselineskip by 100}

\newcommand{\newc}{\newcommand}
\newc{\qbar}{{\overline q}}
\newc{\Kahler}{K\"ahler }
\newc{\deltaGS}{\delta_{\rm GS}}
\usepackage{enumerate}
\usepackage{slashed}
\usepackage{braket}

\usepackage{tikz}

\newcommand{\of}[1]{\!\left( #1 \right)}
\newcommand{\ofp}[1]{\left( #1 \right)}
\newcommand{\sqof}[1]{\left[ #1 \right]}
\newcommand{\cuof}[1]{\left\{ #1 \right\}}

\newcommand{\thetabar}{\overline{\theta}}

\newcommand{\prop}{\Delta}
\newcommand{\epsIR}{\delta}

\newcommand{\coupling}{e}
\begin{document}
\begin{titlepage}
\begin{flushright}
{\large 
SCIPP 16/08\\%UTTG-08-15 \\
}
\end{flushright}

\vskip 1.2cm

\begin{center}

{\large \bf Perturbation Theory in Supersymmetric QED:  Infrared Divergences and Gauge Invariance}

\vskip 1.4cm

{\large   Michael Dine$^{(a)}$, Patrick Draper$^{(b,d)}$, Howard E.~Haber$^{(a,c)}$, \\ Laurel Stephenson Haskins$^{(a)}$}
\\
\vskip 0.4cm
{\it $^{(a)}$Santa Cruz Institute for Particle Physics and
\\ Department of Physics,
     Santa Cruz CA 95064  } \\
\vspace{0.3cm}
{\it
$^{(b)}$Department of Physics, University of California, Santa Barbara, CA 93106
}\\
\vspace{0.3cm}
{\it $^{(c)}$Kavli Institute for Theoretical Physics, University of California,\\ Santa Barbara, CA 93106}\\
\vspace{0.3cm}
{\it $^{(d)}$Amherst Center for Fundamental Interactions, Department of Physics,\\ University of Massachusetts, Amherst, MA 01003
}\\
\vspace{0.3cm}
\vskip 4pt

\vskip 1.5cm

\begin{abstract}
We study some aspects of perturbation theory in $N=1$ supersymmetric abelian
gauge theories with massive charged matter.  In general gauges, infrared (IR) divergences and nonlocal behavior arise in 1PI diagrams, associated with a $1/k^4$ term in the propagator for the vector superfield. 
We examine this structure in supersymmetric QED.  The IR divergences are gauge-dependent and must cancel in physical quantities like the electron pole mass.  We demonstrate that cancellation takes place in a nontrivial way, amounting to a reorganization of the perturbative series from powers of $\coupling^2$ to powers of $\coupling$. We also show how these complications are avoided in cases where a Wilsonian effective action can be defined.

\end{abstract}

\end{center}

\vskip 1.0 cm

\end{titlepage}
\setcounter{footnote}{0} \setcounter{page}{2}
\setcounter{section}{0} \setcounter{subsection}{0}
\setcounter{subsubsection}{0}
\setcounter{figure}{0}

%%%%%%%%%%%%%%%%%%%%%%%%%%%%%%%%%%%%%%%%%%%
%%%%%%%%%%%%%%%%%%%%%%%%%%%%
\singleandthirdspaced

\section{Introduction}
\label{intro}

In weakly coupled supersymmetric field theories, it is convenient for certain applications to employ a manifestly supersymmetric perturbation theory.
For example, non-renormalization theorems were first proven using supergraph techniques~\cite{Grisaru:1979wc}.  These proofs rely on the existence of a particular infrared-safe choice of gauge, analogous at one-loop order to Feynman gauge in non-supersymmetric QED. 

However, in other supersymmetric gauges, perturbation theory is plagued by unphysical infrared divergences. Difficulties may be anticipated from the superspace propagator for the vector superfield. 
In this paper, we focus on the supersymmetric extension of QED~\cite{Wess:1974jb} (henceforth denoted as SQED).
In the supersymmetric $R_\xi$ gauge~\cite{Ovrut:1981wa,Miller:1983pg}, the Lagrangian of SQED  is supplemented by a gauge fixing term:\footnote{Our conventions for supersymmetric notation follow that of Ref.~\cite{Sohnius:1985qm}.}
\begin{align}
\mathscr{L} = & \int d^4 \theta \ \Phi_+^\dagger{}e^{2 \coupling V} \Phi_+{} + \int d^4 \theta \ \Phi_-^\dagger{} e^{-2 \coupling V} \Phi_-{} + \sqof{   \int d^2\theta \ m \Phi_+{} \Phi_-{} + \mathrm{h.c.} }  \nonumber\\
& + \sqof{ {1\over 4} \int d^2\theta \mathcal{W}^\alpha \mathcal{W}_{\alpha}+\mathrm{h.c.} }
- {\xi \over 8} \int d^4 \theta \ofp{ D^2 V} \ofp{ \bar{D}^2 V },
 \label{eq:SUSYQEDLagrangian}
\end{align}
leading to the vector superfield propagator\footnote{Note that the normalization of this propagator differs by a factor of two from the vector superfield propagator given in Ref.~\cite{wessandbagger}.}
\begin{align}
i \Delta_V(k,\theta_1,\theta_2) = {i \over  k^4}(1 - {1 \over \xi}) e^{(\theta_1 \sigma^\mu \bar \theta_2 - \theta_2 \sigma^\mu \bar \theta_1) k_\mu} 
- {i \over 4 k^2} (1 + {1 \over \xi}) \delta^4(\theta_1 -\theta_2) e^{(\theta_1 \sigma^\mu \bar \theta_2 - \theta_2 \sigma^\mu \bar \theta_1) k_\mu}\;.
\label{vpropagator}
\end{align}
It is striking that away from $\xi=1$, the propagator behaves as $1/k^4$ for small $k$. This 
%unusual 
behavior can lead to infrared (IR) divergences in loop graphs that probe the small-$k$ modes of $V$. 

We will exhibit such IR divergences in one-loop contributions to the two-point functions of SQED with massive charged matter. The appearance of infrared issues has been noted in the past.  Ref.~\cite{Abbott:1984pz} described a resolution in
non-abelian gauge theories involving the introduction of a nonlocal gauge fixing term and adjusting the gauge fixing parameter to eliminate the divergences order by order in perturbation theory. In massive abelian theories, this procedure simplifies to the adjustment of the gauge fixing parameter without modifying the gauge fixing term itself.

On the other hand, it is also not difficult to regulate the infrared in a gauge-invariant way. Since the divergences are gauge-dependent, they are unphysical, and must eventually drop out of observable quantities. We study this cancellation in the pole mass of the electron chiral supermultiplet. We find an interesting structure: near the pole, the perturbative series for the two-point functions reorganizes itself. Whereas the na\"ive one-loop graphs contribute to the series at ${\cal O}(\coupling^2)$, near the pole, some -- but not all -- of the graphs exhibit singularities that enhance their contributions to ${\cal O}(\coupling)$. We find exact cancellation between the ${\cal O}(\coupling)$ one-loop graphs, including their IR-divergent pieces. We argue that the cancellation of IR-divergent terms at ${\cal O}(\coupling^2)$ must occur between a combination of the remaining one-loop and enhanced two-loop graphs.

If light neutral fields are added to the theory, the charged massive fields may be integrated out to obtain a Wilsonian effective action subject to the na\"ive non-renormalization theorems. The resulting wave function renormalization should be gauge invariant, and in particular infrared divergences should cancel order by order in the effective action.  We will verify this at low orders in the perturbation series.

This paper is organized as follows.  
In Sec.~\ref{ptreview}, we review and collect the SQED superspace and component propagators relevant for our analysis.
In Sec.~\ref{susyqed}, after recalling the gauge dependence of the mass renormalization in ordinary QED, we describe the one loop renormalization of the electron mass in SQED.  We show that there are infrared divergences and nonlocal behavior in the 1PI corrections to the helicity-flip and helicity-preserving propagators. The nonlocal factors are singular near the mass shell and lead to a mixing of loop orders at fixed order in $e$. We show that the leading IR divergent gauge dependence cancels in the one-loop electron pole mass, while subleading unphysical contributions must cancel against two-loop terms. We also recover the well-known result~\cite{Ferrara:1975ye} that the ultraviolet (UV) divergent part of the mass renormalization is gauge invariant. In Sec.~\ref{wilsonianmodel}, we couple a massless, neutral field to the charged fields and demonstrate cancellation of infrared contributions to the self energy at two loops.  We discuss the implications for the Wilsonian effective action in this case.  In Sec.~\ref{sec:ifpt}, we demonstrate the presence of infrared divergences at higher order in the gauge $\xi=1$, but show that it is possible to choose a gauge, order by order, in which infrared divergences are absent.  In Sec.~\ref{sec:concl} we summarize and conclude.  

Additional background material and further results are collected in three appendices.
In Appendix A, we review mass and wave function renormalization of non-supersymmetric QED.  In Appendix B, 
we discuss the computation of the tree-level propagators of supersymmetric QED and examine the supersymmetric relations among the two-point functions.  Finally, in Appendix C, we demonstrate that the in addition to the divergences, the finite corrections to the physical electron mass also vanish in the on-shell limit. 

\section{Perturbation Theory in SQED }
\label{ptreview}
For convenience, in this section we collect the well-known propagator expressions in SQED.
In superspace, the vector propagator was given in Eq.~(\ref{vpropagator}) and is repeated here for the convenience of the reader,
\begin{align}
i\Delta_V(k,\theta_1,\theta_2) = {i \over  k^4}(1 - {1 \over \xi}) e^{(\theta_1 \sigma^\mu \bar \theta_2 - \theta_2 \sigma^\mu \bar \theta_1) k_\mu} 
- {i \over 4 k^2} (1 + {1 \over \xi}) \delta^4(\theta_1 -\theta_2) e^{(\theta_1 \sigma^\mu \bar \theta_2 - \theta_2 \sigma^\mu \bar \theta_1) k_\mu}\;.
\label{vpropagator2}
\end{align}
The corresponding propagators of the component fields are given in Appendix B.

We study the theory with massive electrons, with the superpotential given by
\beq
W = m \Phi_+ \Phi_-\;,
\eeq
for which the superfield propagators are
\begin{align}
i \prop_{\Phi_{\pm} \Phi_{\mp} } \of{k,\theta_1,\theta_2} &= - i m \delta^{(2)}\of{\theta_1-\theta_2} \exp\sqof{  \ofp{ \theta_1\sigma^\mu\thetabar_1 - \theta_2\sigma^\mu\thetabar_2} k_\mu } \frac{1}{k^2 - m^2} \\
i \prop_{\Phi_{\pm}^\dagger \Phi_{\mp}^\dagger}\of{k,\theta_1,\theta_2} 
 &= + i m \delta^{(2)}\of{\thetabar_1-\thetabar_2} \exp\sqof{  \ofp{ \theta_1\sigma^\mu\thetabar_1 - \theta_2\sigma^\mu\thetabar_2} k_\mu } \frac{1}{k^2 - m^2}\;,\\
i \prop_{\Phi_{\pm}\Phi_{\pm}^\dagger}\of{k,\theta_1,\theta_{2}} &= i \exp\sqof{   \ofp{\theta_1\sigma^\mu \thetabar_1 - \theta_2 \sigma^\mu \thetabar_2 + 2 \theta_1 \sigma^\mu \thetabar_2 } k_\mu }\frac {1}{k^2 - m^2} \;.
\end{align}
It is helpful (and in many computations simpler) to work with a mixture of component and superspace formalisms. We parametrize the vector superfield components as
\begin{align}
V(x,\theta,\thetabar) &= a(x) + i \theta \chi(x) -i \bar \theta \bar \chi(x) + \theta^2 M(x) + \bar \theta^2 \bar M(x) + i \theta \sigma^\mu \bar \theta A_\mu(x)\\
&+ i \theta^2 \bar \theta \left( \bar{\lambda} (x) -\tfrac12 i  \overline{\sigma}^\mu \partial_\mu \chi\of{x} \right) -i \bar \theta^2 \theta \left( \lambda(x) -\tfrac12 i \sigma^\mu\partial_\mu \bar{\chi}\of{x} \right)\nn
+ \tfrac12 \theta^2 \bar \theta^2 \left( D(x)  - \tfrac12 \Box a(x) \right).
\end{align}
In Appendix B.1, we show that the
component Lagrangian for the vector includes the terms,
\beq
{\cal L}_V = \tfrac12 \left( 1- \xi \right) D^2 - \tfrac12 \xi ( \Box a )^2 + \xi D \Box a 
\eeq
 Inverting the quadratic form gives for the momentum space propagators\footnote{Here $\langle DD\rangle$ is defined such that $ \langle 0 | T D\of{x} D\of{y} | 0 \rangle_{\mathrm{F.T.}} \equiv (2\pi)^{-4}\int d^4 k \,   \langle D D \rangle \exp\bigl[ - i k\newcdot(x-y)\bigr]$. Similar expressions  apply to $\langle a D \rangle$ and $\langle a a \rangle$.} of $a$ and $D$:
 \begin{align}
 \langle D D \rangle &= i, \\
 \langle a D \rangle &=-\, \frac{i}{k^2},\\
 \langle a a \rangle &= \left(1 - {1 \over \xi}\right) \frac{i}{k^4} .
 \label{componentpropagators}
 \end{align}
 We see that in components, the $1/k^4$ infrared behavior discussed in Sec.~\ref{intro} can be traced to the kinetic term for the lowest component of $V$, which contains four derivatives in a general gauge.
 In the $\xi =1$ gauge, the $1/k^4$ terms disappear, and severe infrared divergences are avoided in low orders of perturbation theory.
However, loop corrections will  reintroduce $1/k^4$ terms in the propagator.  In particular, at the level of component fields, there is an $\langle a D \rangle$ propagator, and at one loop charged fields correct the $\langle D D\rangle$ two point function at zero momentum.

\section{Self-Energies in SQED}
\label{susyqed}

The na\"ive expectation from the non-renormalization theorems is that there should be no renormalization of the superpotential mass $m$ arising from the $\Phi_+ \Phi_-$ self-energy.  Any renormalization of the physical mass should arise as a result of corrections to the \Kahler potential.

In supersymmetric Feynman gauge, $\xi=1$, there are indeed no one-loop 1PI contributions to $\langle\Phi_+ \Phi_-\rangle$.  This can be seen directly in superspace, as in Ref.~\cite{Grisaru:1979wc}.  It can also be seen by working in components with explicit auxiliary fields. We take as the component expansion of the chiral superfields:
\begin{align}
\Phi_{\pm}(x,\theta,\thetabar)=\exp(-i\theta\sigma^\mu\thetabar\partial_\mu)\bigl[\phi_{\pm}\of{x} + \sqrt{2} \theta \psi_{\pm}\of{x}+ \theta\theta F_{\pm}\of{x}\bigr]\,.
\end{align}
In particular, such a two point function for the superfields would yield, in components, a non-vanishing $\langle F_+ \phi_- + F_- \phi_+\rangle$.  But it is easy to see there is no such graph at one loop.    There is a wave function renormalization for $\Phi_+$ and $\Phi_-$ which is ultraviolet divergent and corrects the physical mass.  

In more general gauges, the situation is more complicated.  At one loop, there are UV-finite, IR-divergent, nonlocal contributions to  $\langle\Phi^+ \Phi^-\rangle$. The apparent violation of nonrenormalization is of the form discussed in Refs.~\cite{Howe:1989az,West:1990rm,Jack:1990pd} and attributable to the nonlocal nature of 1PI effective actions~\cite{Seiberg:1993vc}. There are also  corrections to $\langle\Phi^\pm \Phi^{\pm\dagger}\rangle$ that are both UV {\it and} IR
divergent.  Only suitable physical questions are expected to yield finite and
gauge-invariant answers.  
The new feature for $\xi\neq 1$, namely the infrared divergences, arise from the $1/k^4$ term in the vector superfield propagator noted above.

To see these divergences explicitly, it is convenient to focus on two types of self-energies involving the scalar components of the electron supermultiplets: $\braket{F_+ \phi_-}$ corresponding to a helicity flip process, and $\braket{ F_+^* F_+ }$ corresponding to a helicity preserving process.

\begin{figure}[t!]
\begin{center}
\includegraphics[width=.3\linewidth]{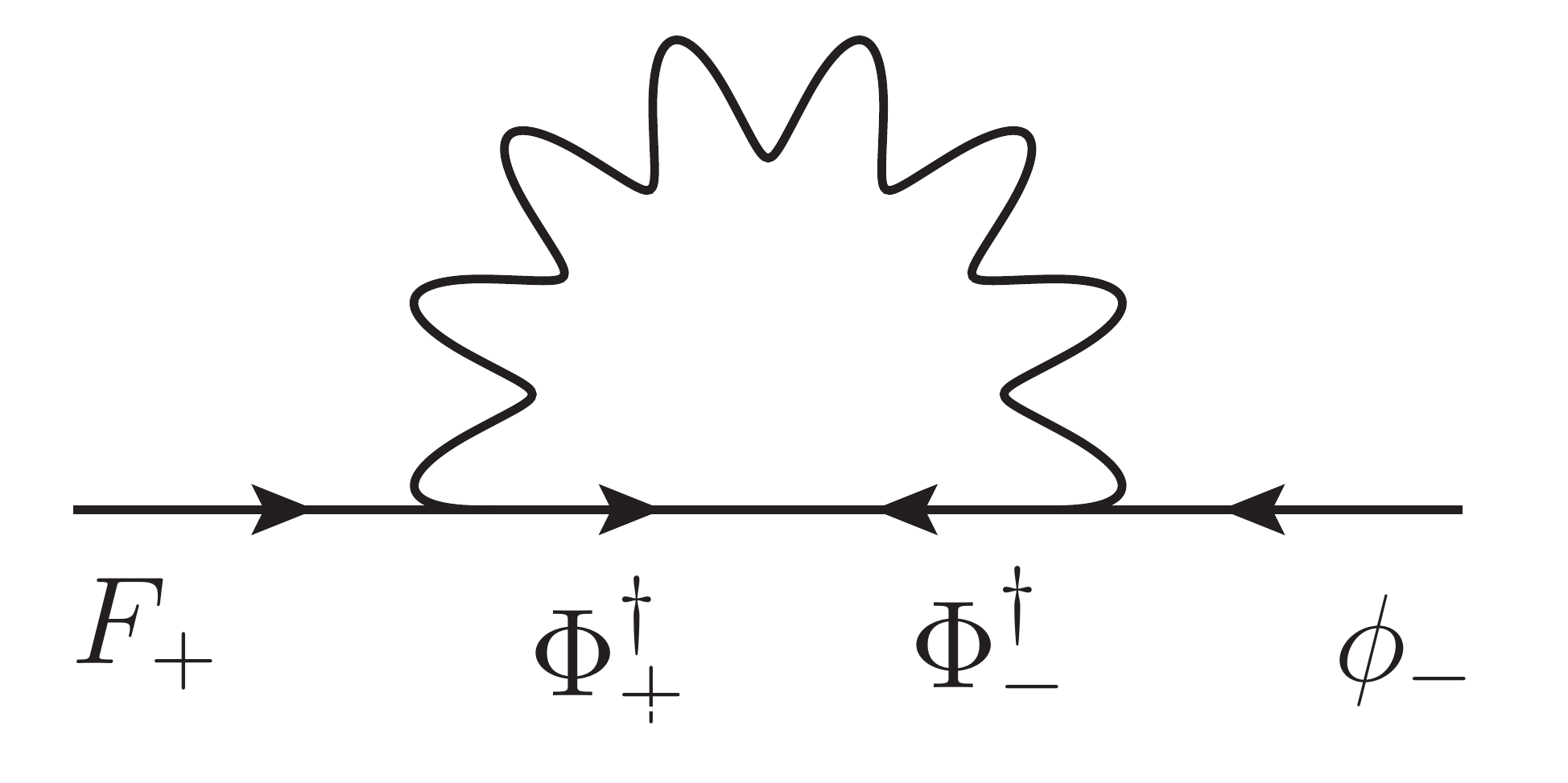}
\caption{One loop contribution to helicity flip process.}
\label{fig:helicityflip}
\end{center}
\end{figure}

Corrections to the $\braket{F_+ \phi_-}$ propagator come from the diagram shown in Fig.~\ref{fig:helicityflip}.
In terms of the component fields, only $a$ couples to $F^\dagger F$ and propagates along the vector line. We obtain: 
\begin{align}
I_{F_+ \phi_- }   = 
  -  \coupling^2   m \ofp{1-{1\over \xi}} p^2 \int { d^4 k \over \ofp{2\pi}^4 } {1\over k^4 \sqof{  \ofp{ p-k}^2 - m^2 } }\;,
\label{eq:1scalar_integral_form}
\end{align}
which is IR-divergent and UV-finite.
Focusing on the small $k$ region yields:
\begin{align}
\left(I_{F_+ \phi_-}\right)_\mathrm{IR}  \sim  - \coupling^2 m \ofp{ 1 - {1\over \xi}} { p^2 \over p^2 - m^2 } \int {d^4 k \over \ofp{ 2 \pi}^4 } {1\over k^4} \ \ \ \ \mathrm{for\ } k^2 \ll p^2 .
\end{align}
At one loop we can cut off the infrared divergence at a small momentum ``by hand," or by introducing a small mass for the 
vector superfield.  Dimensional regularization~\cite{'tHooft:1972fi,Leibbrandt:1975dj} with $d =   4 - 2 \delta $ and $\epsIR <0$ provides a gauge-invariant IR regulator~\cite{Marciano:1975de}. The IR divergent part is
\begin{align}
\left(I_{F_+ \phi_-}\right)_\mathrm{IR} = - i  m  {\coupling^2 \over 16 \pi^2 }   \ofp{ 1 - {1\over \xi}} { p^2 \over p^2 - m^2 } { 1 \over \epsIR }\;.
\label{fphiirdivergence}
\end{align}

\begin{figure}[b!]
\begin{center}
%\vspace{1cm}
%\begin{fmffile}{sunset}
%      \setlength{\unitlength}{1cm}
%      \begin{fmfgraph*}(4.5,3.5)
%        \fmfleft{i1}
%        \fmfright{o1}
%        \fmflabel{$F_+^*$}{i1}
%         \fmflabel{$F_+$}{o1}
%        \fmf{plain_arrow,tension=3}{i1,v1}
%        \fmfset{wiggly_len}{10}
%         \fmf{wiggly,left=1,tension=0.5}{v1,v2}
%                  \fmf{plain_arrow}{v1,v2}
%        \fmf{plain_arrow,tension=3}{v2,o1}
%      \end{fmfgraph*}
%    \end{fmffile}
%\hspace{2cm}
%\begin{fmffile}{seagull}
%      \setlength{\unitlength}{1cm}
%      \begin{fmfgraph*}(4.5,3.5)
%        \fmfleft{i1}
%        \fmfright{o1}
%        \fmftop{p1}
%        \fmfbottom{p2}
%        \fmflabel{$F_+^*$}{i1}
%         \fmflabel{$F_+$}{o1}
%        \fmf{plain_arrow}{i1,v1}
%        \fmf{phantom,tension=1}{p1,v2}
%        \fmf{phantom,tension=.3}{p2,v1}
%                \fmfset{wiggly_len}{10}
%         \fmf{wiggly,left=1,tension=.2}{v1,v2,v1}
%                 \fmf{plain_arrow}{v1,o1}
%      \end{fmfgraph*}
%    \end{fmffile}
%
\includegraphics[width=\linewidth]{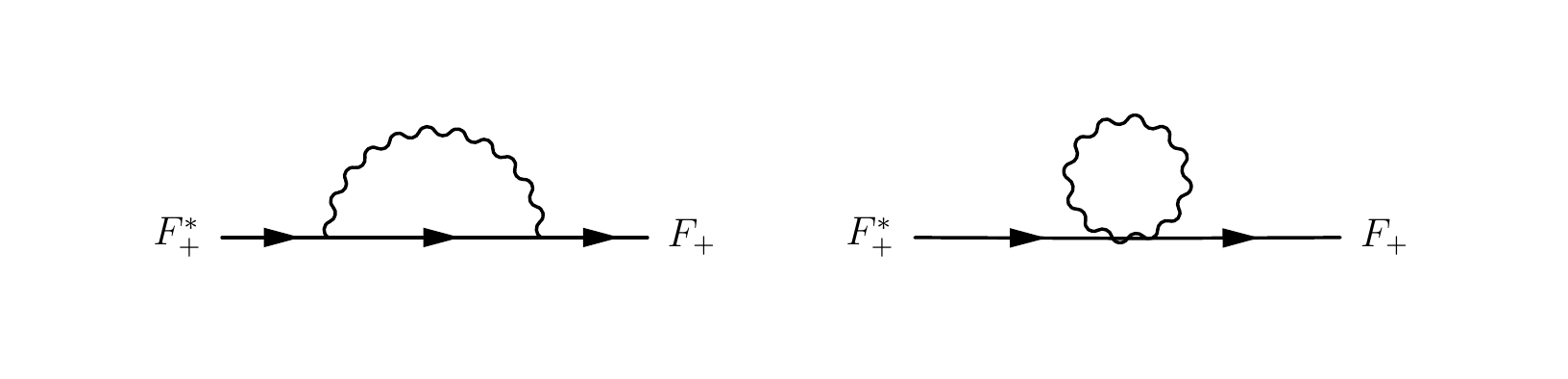}
\caption{One loop contributions to the helicity preserving process.}
\label{wavefunctioncorrections}
\end{center}
 \end{figure}
 
The $\braket{ F_+^* F_+ }$ propagator receives corrections from the ``sunset'' and ``seagull'' diagrams shown in Fig.~\ref{wavefunctioncorrections}.
Both the $a$ and $M$ components of $V$ propagate in the sunset diagram. The former gives rise to an IR singularity, while the latter provides a ultraviolet divergence. The Feynman integrals contributing to the sunset diagram are,
\begin{align}
\begin{split}
I_{F_+^* F_+}^{\rm sun} 
=  -{ \coupling^2 \over 2}
\cuof{
  -2  \ofp{ 1 - {1\over \xi } } \int {d^4 k \over \ofp{  2 \pi}^4 } {  p^2 - p\cdot k \over k^4 \sqof{ \ofp{p-k}^2 - m^2 } }  + {1\over \xi}  \int {d^4 k \over\ofp{2\pi}^4 } { 1 \over k^2 \sqof{ \ofp{p-k}^2 - m^2 } }
  } . 
  \end{split}  \label{eq:2scalar_integral_form}
    \end{align}
We isolate the IR divergence in the first integral with dimensional regularization by integrating over $d = 4 - 2\delta$ dimensions, where $\delta<0$,
 \begin{align}
\left(I_{F_+^* F_+}^{\rm sun}\right)_\mathrm{IR} = - i {  \coupling^2 \over 16 \pi^2 } \ofp{ 1 - {1\over \xi}} { p^2 \over p^2 - m^2 } {1\over \epsIR}.
\label{fstarfirdivergence}
\end{align}
The ultraviolet divergent part, the second integral in~\ref{eq:2scalar_integral_form}, may also be isolated with dimensional regularization~\cite{Siegel:1979wq,Capper:1979ns,Jack:1997sr},  by taking $d = 4 - 2\epsilon$ with $\epsilon>0$,
\begin{align}
\left(I^{\rm sun}_{F_+^* F_+}\right)_\mathrm{UV} = - {i \over 2}  {\coupling^2 \over    16 \pi^2 } {1\over \xi} { 1 \over \epsilon }.
\label{eq:uv1}
\end{align}

The $a$ component of $V$ propagates in the seagull loop, giving
\beq
 I_{F_+^* F_+ }^{\rm sea}  = {- \coupling^2 \over 2}  \ofp{  1 - {1\over \xi} } \int {d^4 k \over \ofp{ 2\pi}^4 }{ 1 \over k^4 } \;.
\eeq
This Feynman integral is both UV and IR divergent. 
Scaleless integrals may be consistently set to zero in dimensional regularization~\cite{Leibbrandt:1975dj}, so it is sometimes said that the UV and IR divergences cancel. Ultimately this property will be unimportant for our analysis. Moreover, we would like to keep these divergences separate at one loop, so we retain the $\epsilon$, $\delta$ notation of Eqs.~(\ref{eq:3scalarUV}) and (\ref{eq:3scalarIR}) to keep the origin of the  divergences distinct.

The UV and IR divergent pieces are:
\begin{align}
\left(I_{F_+^* F_+}^{\rm sea}\right)_\mathrm{UV} = - {i\over 2 }{ \coupling^2 \over 16 \pi^2 }  \ofp{  1- {1\over \xi}} {1\over \epsilon}\;,
\label{eq:3scalarUV}
\end{align}
\begin{align}
\left(I_{F_+^* F_+}^{\rm sea}\right)_\mathrm{IR} = + { i \over 2} { \coupling^2 \over 16 \pi^2 } \ofp{  1 - {1\over \xi}} {1\over \epsIR }\;.
\label{eq:3scalarIR}
\end{align}

We have seen that the $\Phi_+ \Phi_-$ propagator is UV finite in any $R_\xi$ gauge. Therefore, if the physical mass is to
be gauge invariant, the ultraviolet divergent pieces of the wave function renormalization must be gauge invariant~\cite{Ferrara:1975ye}.
This property is manifest in the sum of Eqs.~(\ref{eq:uv1}) and (\ref{eq:3scalarUV}), where terms
proportional to $(1/\epsilon)\cdot(1 /\xi)$ cancel.

\section{The Electron Pole Mass}
\label{polemass}

In Appendix \ref{ordinaryqed}, we review the one-loop correction to the mass of the electron in ordinary QED in the $R_\xi$ gauges. In brief, the quadratic terms in the bare 1PI effective action have the form
\begin{align}
{\cal L}\ni \left(1+a(p)\right)\bar\psi_0\pslash \psi_0 - m_0 \left(1+b(p)\right) \bar\psi_0\psi_0+\dots\;.
\end{align}
In canonical normalization, one can define a one-loop ``mass shift" for general $p$, 
\begin{align}
\delta m(p) = m_0\bigl[b(p)-a(p)\bigr]\;.
\end{align}
The ultraviolet divergence in the mass shift (giving rise to the $\beta$-function for the renormalized mass parameter) is gauge invariant, with gauge-dependent terms canceling between the helicity-flip and helicity-preserving contributions to the self-energy. However, in a given renormalization scheme the finite pieces of the mass shift are only gauge
invariant on-shell.  In the case of SQED, we might expect something similar, with gauge invariance -- and now the cancellation
of infrared divergences -- holding only on-shell.

There is an extra subtlety in SQED due to the non-local, $(p^2 - m^2)^{-1}$ behavior we have seen in general $R_\xi$
gauges. 
If the one-loop pole mass is shifted from the tree-level mass by a power of $\coupling^2$, as in most renormalization schemes, factors of $(p^2-m^2)^{-1}$ can spoil the na\"ive ordering of loop corrections in powers of $\coupling^2$.
(This is analogous to issues with ordering in $e$ in finite temperature perturbation theory
in ordinary gauge theories.)  In the next two subsections, we examine this subtlety and the gauge invariance of the supersymmetric electron mass in greater detail.

\subsection{Nonlocality and the SQED loop expansion}
\label{perturbationtheorystructure}

The quadratic terms in the renormalized effective action involving the scalar component fields $\phi_\pm$ and $F_\pm$ may be written as
\begin{align}
\mathscr{L}^{\rm eff}_{\phi F} = 
&  \ofp{ \begin{array}{cc} F_+^*  & \phi_-  \end{array} } 
\Delta_{F\phi}^{-1}
\ofp{ \begin{array}{c}  F_+ \\ \phi_-^*  \end{array} }  + \ofp{ +\leftrightarrow -  }\,,
\end{align}
where $\Delta_{F\phi}^{-1} $ is the inverse propagator matrix [cf.~\eq{dfi}].  In momentum space, we can write
\begin{align}
\Delta_{F\phi}^{-1} \equiv  \ofp{ \begin{array}{cc} 1+ A\of{p}  &  \quad m\bigl[1 + B\of{p}\big]  \\[6pt]
m\bigl[1 + B\of{p}\bigr] & \quad p^2\bigl[1 + A\of{p}\bigr] \end{array} },
\label{eq:scalarMM}
\end{align}
where $A$ and $B$ are proportional to $\overline{\rm DR}$-renormalized self-energies~\cite{Siegel:1979wq,Capper:1979ns,Martin:2005ch}, the IR divergent pieces of which were computed above.
The pole mass for the multiplet is determined by solving $\det\ofp{  \Delta_{F\phi}^{-1} } = 0$,
\begin{align}
p^2-m^2 = 2F(p) m^2\;,\;\;\;\;F(p)\equiv\frac{1}{2}\left[\left(\frac{1+B(p)}{1+A(p)}\right)^2-1\right]\;.
\label{eq:polemasseq}
\end{align}
$F$ admits an expansion in powers of $\coupling^2$. At ${\cal O}(\coupling^2)$, $F=(B-A)$, and $A$ and $B$ correspond to the UV-subtracted one-loop diagrams of Sec.~\ref{susyqed}. Thus, if
$F(p)$
is well-behaved near $p^2 = m^2$, the physical mass receives a one-loop correction of ${\cal O}(\coupling^2)$, $m_{\rm phys}-m=mF(m)$. 

However, if $A$ or $B$ have singularities associated with nonlocal terms, the link between loops and powers of $\coupling^2$ can break down. For a simple toy example at one-loop order, take the following form for $F(p)$:
\begin{align}
F(p) = \coupling^2 \left(f_1^s{p^2 \over p^2 - m^2}+f_1^n\right) \;.
\label{eq:Fsing}
\end{align}
Here we have included a singular piece with constant coefficient $f_1^s$ and a nonsingular piece with coefficient $f_1^n$.

Then
the leading correction to the mass is
\begin{align}
m_{\rm phys}-m=\pm em\sqrt{f^s_1/2}+{\cal O}(\coupling^2) \;.
\label{eq:oneloopsingular}
\end{align}
We see that the singular term contributes to the mass with one less power of $e$ than the nonsingular term. Likewise, it is easy to see that two-loop contributions to $F(p)$ proportional to the same nonlocal singularity can contribute at ${\cal O}(\coupling^2)$, the same as one-loop nonsingular terms.

The ambiguity in the sign in Eq.~(\ref{eq:oneloopsingular}) can only be resolved if cancellations between $A(p)$ and $B(p)$ are such that $f_1^s=0$, in which case the mass is not actually corrected at ${\cal O}(e)$. Indeed, singularities like those in this toy example appeared in the computation above of the supersymmetric electron self-energies, and in SQED we expect the ${\cal O}(e)$ terms in the electron pole mass to cancel for other reasons:  the singularities are associated with unphysical, gauge-dependent, IR-divergent terms, and ${\cal O}(e)$ corrections are not present in Feynman gauge. The lessons we learn are:
\begin{enumerate}
\item{The leading-order cancellation will take place only between one-loop graphs with nonlocal singularities.}
\item{At higher orders in $e$, singularities must
cancel between different loop orders.}
\end{enumerate}

\subsection{Cancellation of ${\cal O}(e)$ terms in $m_{phys}$}

In the previous subsection, we saw that the appearance of nonlocal singularities in the supersymmetric electron self-energies, combined with the requirement of gauge invariance, implies the existence of cancellations between contributions at different loop orders. In Appendix~\ref{finitecorrections}, we verify the exact cancellation for the terms at ${\cal O}(e)$, arising from the helicity flip diagram and the helicity preserving sunset diagram, both of which have singularities as $p^2$ goes on-shell. Here, for brevity, we show only the cancellation of the IR divergent pieces at ${\cal O}(e)$ arising from those diagrams.

From the results of Sec.~\ref{susyqed}, we have
\begin{align}
A\of{p} =& - i { \coupling^2 \over 16 \pi^2 } \ofp{ 1 - {1\over \xi}} {  p^2 \over p^2 - m^2 } {1\over \epsIR } + \mathrm{finite} \\
B\of{p} =& - i {\coupling^2 \over 16 \pi^2 } \ofp { 1 - {1\over \xi}}  { p^2 \over p^2 - m^2 } { 1 \over \epsIR} + \mathrm{finite}\;,
\end{align}
where $A$ and $B$ are defined in Eq.~(\ref{eq:scalarMM}).
Consistent with our discussion in the previous subsection, we have neglected the  seagull diagram in $B$. The seagull contributions are nonsingular and contribute to the pole mass only at ${\cal O}(\coupling^2)$. We see that the gauge-dependent IR divergences cancel in the combination $B-A$ appearing in the pole mass.

Although it enters at ${\cal O}(\coupling^2)$, there
is an unphysical IR divergence in the seagull diagram, which must be cancelled by a two-loop
contribution to $F(p^2)$ proportional to $g^4 (p^2 - m^2)^{-1}$. At two-loop order there are also double IR divergences associated with graphs with two vector superfield propagators. We expect the complete structure of cancellations to be quite intricate.

 %%%
  
\section{Integrating Out Massive Charged Fields}
\label{wilsonianmodel}

The real power of the non-renormalization theorems arises in situations where a Wilsonian effective action is useful. It is interesting to see how the gauge artifacts discussed above, and in particular the infrared divergences for $\xi\neq 1$, cancel when massive fields are integrated out to obtain
a low energy effective action for a set of light fields.

A simple example is generated by adding a light neutral field to the massive SQED theory, with superpotential
\begin{align}
W = m \Phi_+ \Phi_- + \lambda \Phi_0 \Phi_+ \Phi_- + \lambda^\prime \Phi_0^3\;.
\end{align}
Integrating out the massive $\Phi_\pm$,  we obtain an effective action for $\Phi_0$.  The standard non-renormalization
theorem analysis here would indicate that the only corrections to $\lambda$ 
arise from wave function renormalization.  In this theory, it is easy to check that there are no low order corrections to the 
1PI $\Phi_0^3$ three point function (this can be done with supergraphs, or in components, looking for an $F_0 A_0 A_0$
1PI Green's function).  This is a consequence of a holomorphy-type argument~\cite{Seiberg:1993vc}, treating $\lambda$ as a spurion and
assigning it an $R$ charge.

\begin{figure}[h!]
\begin{center}
\includegraphics[width=\linewidth]{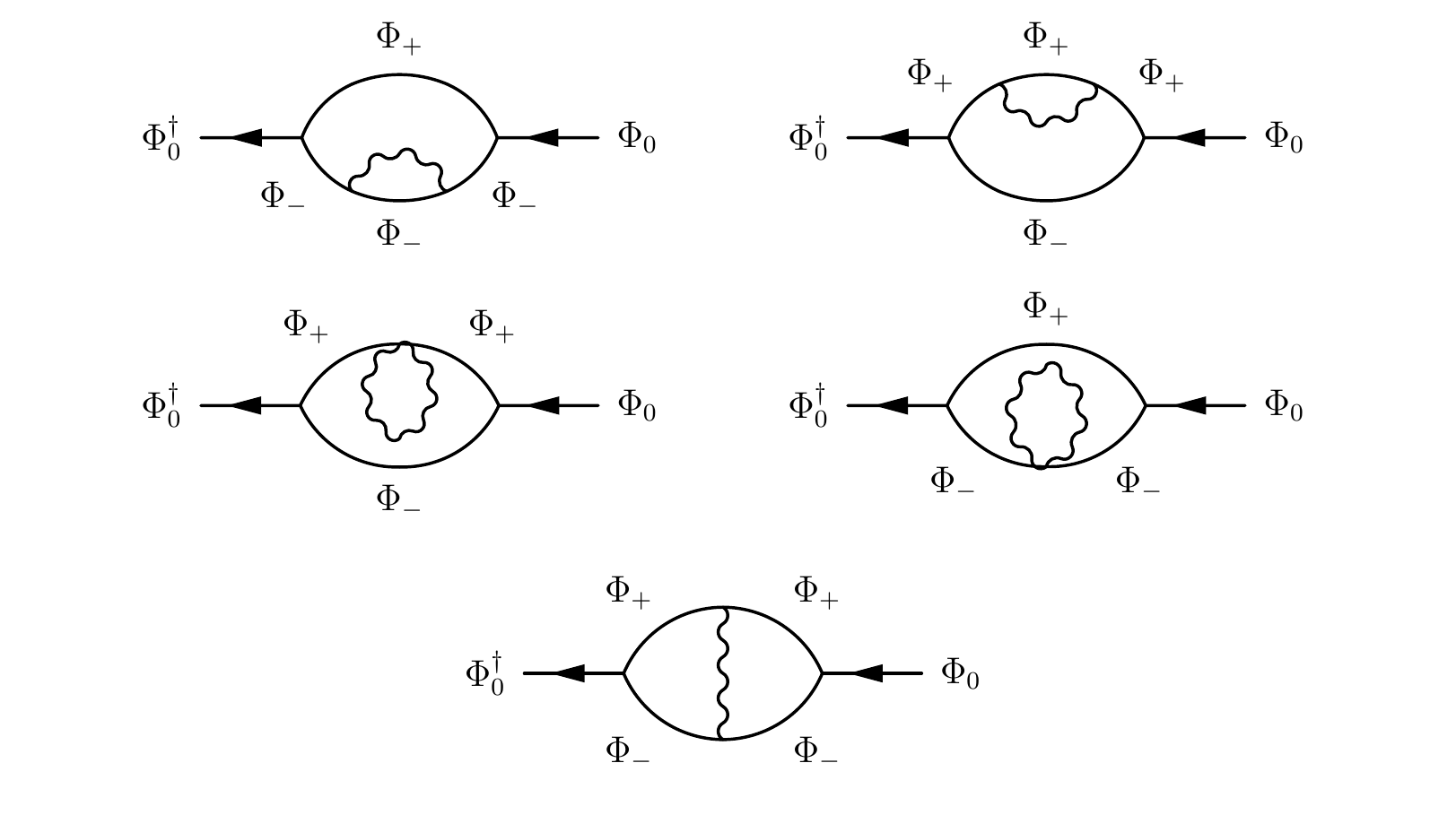}
\caption{Diagrams contributing to the $\Phi_0$ effective lagrangian at two loops.}
\label{heavylight}
\end{center}
\end{figure}

There {\it should} be a renormalization of $\lambda$ proportional to the wave function renormalization of $\Phi_0$.  It should be gauge invariant,
and free of infrared divergences and other pathologies.  Gauge fields enter the wave function renormalization
at two loops.  While the full two-loop computation is complicated, the leading infrared divergent pieces of individual Feynman diagrams
are easily isolated.   There are many diagrams, but only a small set which are both infrared and ultraviolet divergent, and we examine these for illustration.  In particular, diagrams which include helicity flip (i.e. $\langle\Phi_+ \Phi_-\rangle$) propagators are ultraviolet finite, as they come with an positive power of $m$.  This leaves five diagrams, shown in Fig.~\ref{heavylight}.

To see the cancellation of the gauge-dependent pieces, we work at zero external momentum and isolate the leading infrared and ultraviolet divergent piece.
Consider the first of these diagrams.   
It is particularly simple to compute the $F_0 F_0^\dagger$ component propagator.  The diagram is given by (we can now safely Wick rotate to Euclidean space)
\beq
\int {d^4 p \over (2 \pi^4)}  {d^4 k \over (2 \pi^4)} {p^2 \over (p^2 +m^2)^3} {p^2 \over ((p+k)^2 + m^2)} {1 \over k^4}
\eeq
The most singular part of this diagram in the infrared behaves as:
\beq
\int {d^4 p \over (2 \pi^4)}  {d^4 k \over (2 \pi^4)}{1 \over k^4} {p^4 \over (p^2 +m^2)^4}.
\eeq 
This expression diverges for small $k$ and the $k$ integral should be thought of as cut off at~$\vert p \vert$.  The remaining integral over $p$ is UV divergent.  For large $p$, the integral takes the form
\beq
 \int^{\vert \Lambda \vert}_m {d^4 p \over p^4}\int^{\vert p \vert} {d^4 k\over k^4}.
 \eeq
In the limit of small $k$ and large $\Lambda$, all of the integrals take this form, up to constants.

To see the cancellation, then, we need only to determine the relative weights of these diagrams.  The first three diagrams have the same overall weight, but the third has sign opposite to the first two due to the opposite charges of $\Phi_+$ and $\Phi_-$.  The fourth and fifth diagram contain an extra factor of $1/2$, arising from the
expansion of the exponential in $e^{2eV}$ to second order, and an extra minus sign because there is one less propagator and one less vertex.
As a result, the sum is of the form $1+1-1-1/2-1/2=0$, and the leading IR divergences cancel in the effective action.

\section{Infrared Finite Perturbation Theory}
\label{sec:ifpt}
\begin{figure}[h!]
\begin{center}
%\begin{fmffile}{aa}
%      \setlength{\unitlength}{1cm}
%      \begin{fmfgraph*}(4.5,3.5)
%        \fmfleft{i1}
%        \fmfright{o1}
%        \fmf{dots,label={$a \ \ \ D$},tension=2}{i1,v1}
%         \fmf{scalar,label=$\phi_{\pm}^*  \ \ \ \ \ \ \ \ \ \ \phi_{\pm}$,left=1,tension=0.5}{v1,v2}
%         \fmf{scalar,label=$\phi_{\pm}      \ \ \ \ \ \ \ \ \ \ \phi_{\pm}^*$,left=1,tension=0.5}{v2,v1}
%        \fmf{dots,label={$D \ \ \ a$},tension=2}{v2,o1}
%      \end{fmfgraph*}
%    \end{fmffile}
\includegraphics[width=\linewidth]{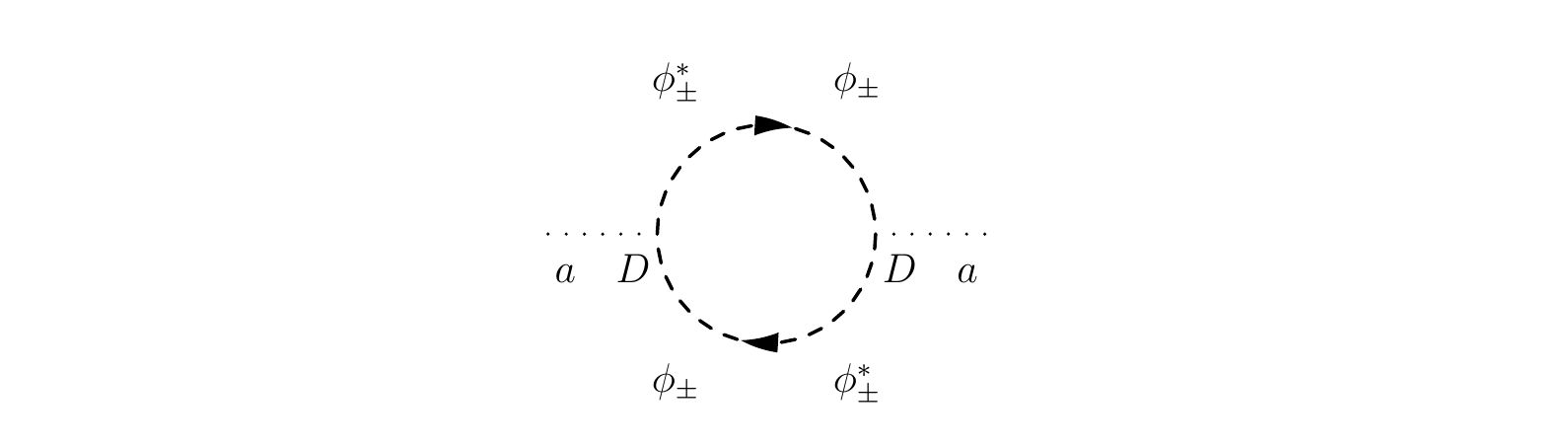}
\caption{Correction to the $\langle aa \rangle$ two point function.}
\label{fig:aacorr}
\end{center}
\end{figure}

IR divergences arise from the lowest component of the vector superfield. We could avoid the whole issue of IR divergences at one loop by choosing $\xi = 1$ for our computations, as in Ref.~\cite{Grisaru:1979wc}.
However, as noted in Ref.~\cite{Abbott:1984pz}, even working in Feynman gauge, infrared divergences are still encountered at higher order.
In terms of component fields, the problem is that with $\xi=1$ there still an $\langle a D \rangle$ propagator, proportional to $1/k^2$.
The 1PI $\langle D D\rangle$ two point function is non-vanishing (and UV divergent) at zero momentum, and together with $\langle a D\rangle$, gives rise to
a one-loop $1/k^4$ propagator for $a$ through diagrams like Fig.~\ref{fig:aacorr}.
\noindent This reintroduction of $1/k^4$ can be dealt with by adjusting the gauge condition order by order to cancel it off. For example, at one loop, the $\langle a a \rangle$ propagator becomes
\begin{align}
\delta \langle aa\rangle \;=\; \coupling^2 {1\over 2} \ofp{- {  2 \over k^2 } {1\over \xi } }^2  \int {d^4 q \over \ofp{2\pi}^4 } {  i \over q^2 - m^2 } {  i \over \ofp{ k - q }^2 - m^2 } \;=\;  {8 \over k^4} { 1 \over \xi^2 }  { \coupling^2 \over 16 \pi^2 } {1\over \epsilon }+\dots\;.
\end{align}
After renormalization the $1/\epsilon$ is replaced by $\log(\mu/m)$ at small $k$.

Letting $\xi = 1 + \delta \xi$, we can solve for $\delta \xi$ such that the $1/k^4$ term in the corrected $\braket{ a a }$ propagator vanishes:
\begin{align}
\delta \xi = 8 \coupling^2 {1 \over \ofp{ 4 \pi}^2 }{\log(\mu/m)} + \mathcal{O}\of{g^4 }\;.
\end{align}
This procedure may be iterated order by order in $e$.

Ref.~\cite{Abbott:1984pz} considered nonabelian theories, in which some one loop corrections to the vacuum polarization are always proportional to $\log(k^2)/k^4$ for small external momentum $k$. This new $k$ behavior caused further complications in the removal of IR divergences, in particular necessitating the introduction of new non-local gauge fixing terms. In the massive abelian theory, all logarithms are cut in the infrared by $m$, corresponding to the decoupling of all charged matter and the IR-freedom of the gauge coupling. Therefore we have only to cancel the $1/k^4$ behavior, as above.

\section{Conclusions}
\label{sec:concl}
It is not surprising that there are infrared issues in perturbation theory in supersymmetric gauge theories in general gauges.  From dimensional
analysis alone, it follows that
\beq
\langle a ~a \rangle = {C \over k^4}
\eeq
so to avoid infrared divergences, it is necessary that the $\langle aa\rangle$ two-point function vanish.  This degree of freedom is a gauge artifact;
from the simple existence of Wess-Zumino gauge~\cite{Wess:1974jb}, one expects that there can be no physical effect.

Still, we rely on covariant gauges, and particularly on the manifest supersymmetry, locality (and infrared finiteness) of the effective action, to make important statements, including proofs of non-renormalization theorems.  We have seen here that in situations in which one can integrate out massive fields, so as to obtain a Wilsonian action for light fields, infrared divergences and non-locality cancel.  In discussions of 
1PI actions, it is important to consider physical questions, like the pole masses of stable particles.  We have also explained how one may choose a gauge,
order by order, so that infrared divergences cancel.

\section*{Acknowledgements}

This work was initially inspired by a question on supersymmetric gauge theories raised by John Terning.  Subsequent conversations with John Terning, Stephen Martin and Tim Jones are gratefully acknowledged.   
P.D. is supported in part by the National Science Foundation Grant No.\ PHY13-16748.  M.D., H.E.H. and L.S.H. are supported in part by U.S. Department of Energy grant DE-FG02-04ER41286. 
Some aspects of this work were carried out at 
the Kavli Institute for Theoretical Physics in Santa Barbara, CA, and supported in part by the National Science Foundation under Grant No. NSF PHY11-25915. 

%APPENDIX

\appendix

\section{Gauge Dependence, Wave Function and Mass Renormalization in Non-supersymmetric QED}
\label{ordinaryqed}
In this appendix we review how gauge dependence appears in the electron mass renormalization in non-supersymmetric QED.

In covariant gauges , where the photon propagator is given by\footnote{To make contact with the notation of  \eq{eq:SUSYQEDLagrangian}, we note that $\ahat\equiv\xi^{-1}$.  In this Appendix, we prefer to employ the gauge parameter $\ahat$ in order to follow the standard textbook notation employed in the treatment of QED field theory~\cite{PS}.}
 \beq
 D_{\mu \nu}  = -{1 \over k^2} \left (g_{\mu \nu} - {k_\mu k_\nu \over k^2} (1-\ahat) \right ),
 \eeq 
we compute the 1PI electron two-point function in momentum space:
\beq \label{gamtwo}
i\Gamma^{(2)}(p)=i(\slashed{p}-m)-i\Sigma(p)\,,
\eeq
where $p$ is the four-momentum of the electron.  Here we denote the sum of the loop contributions to $i\Gamma^{(2)}(p)$ by $-i\Sigma(p)$.   
At one-loop, the two contributing Feynman graphs are
\begin{center}
\includegraphics[width=\linewidth]{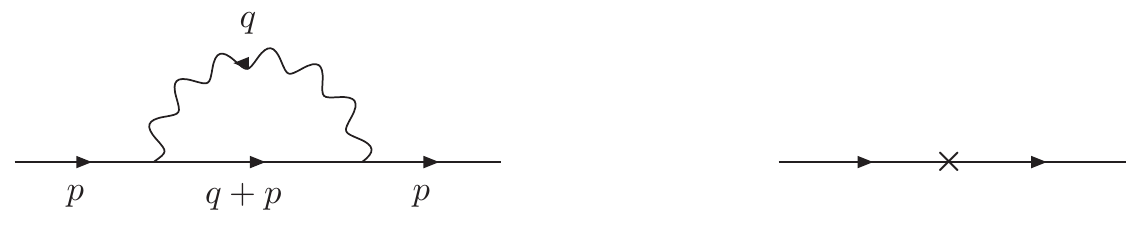}
\end{center}
\vskip -0.15in
\noindent
The cross indicates the contribution of the terms $i\delta Z_2\overline{\psi}\slashed{\partial}\psi
-(\delta Z_m+\delta Z_2)m\overline{\psi}\psi$ of the counterterm Lagrangian,
where $\delta Z_m$ and $\delta Z_2$ are defined such that
\beq
\psi = (1 + \delta Z_{2})^{-1/2} \psi_B, \ \ \ \ m = (1 + \delta Z_m)^{-1} m_B\,,
\eeq
(with subscript $B$ denoting bare quantities and absence thereof denoting renormalized quantities).
At one loop,
\beqa
-i\Sigma(p)&=&(i\mu^\epsilon e)^2\int \frac{d^n q}{(2\pi)^n}\frac{\gamma^\nu(\slashed{q}+\slashed{p}+m)\gamma^\mu}{q^2\bigl[(q+p)^2-m^2\bigr]}\left(g_{\mu\nu}-(1-\ahat)\frac{q_\mu q_\nu}{q^2}\right)
+i\delta Z_2\slashed{p}-im(\delta Z_m+\delta Z_2)\,.\nonumber \\
&&\phantom{line}
\eeqa
Performing the integrals, we obtain
 \beq\label{AB}
 \Sigma(p) = -\pslash A(p)  + m B(p)\;,
 \eeq
 where
\begin{align}
A(p^2)&=\delta Z_2+\frac{\alpha\,\ahat}{4\pi}(4\pi)^{\epsilon}\Gamma(\epsilon) \nonumber\\
&\qquad +\frac{\alpha\,\ahat}{4\pi}\left\{\left(1+\frac{m^2}{p^2}\right)\left[1-\left(1-\frac{m^2}{p^2}\right)\ln\left(1-\frac{p^2}{m^2}\right)\right]-\ln\left(\frac{m^2}{\mu^2}\right)\right\}+\mathcal{O}(\epsilon)\label{eq:Ageneral} \\
B(p^2) &=\delta Z_m+\delta Z_2+\frac{\alpha}{4\pi}(3+\ahat)(4\pi)^{\epsilon}\,\Gamma(\epsilon)
\nonumber
 \\
 &\qquad+\frac{\alpha}{2\pi}\left\{2+\ahat-{1\over 2}(3+\ahat)\left[\left(1-\frac{m^2}{p^2}\right)\ln\left(1-\frac{p^2}{m^2}\right)+\ln\left(\frac{m^2}{\mu^2}\right)\right]\right\}+\mathcal{O}(\epsilon)\;,
\label{eq:Bgeneral}
\end{align}
and $\alpha\equiv e^2/(4\pi)$.

 \subsection{$\overline{\rm MS}$ Renormalization}
In the modified minimal subtraction scheme~\cite{'tHooft:1972fi,Bardeen:1978yd}, the counterterms are 
\beq \label{zees}
\delta Z_2^{\overline{\rm MS}} =-\frac{\alpha\,\ahat}{4\pi}(4\pi)^{\epsilon}\,\Gamma(\epsilon)\,,\qquad\quad
\delta Z_m^{\overline{\rm MS}} =-\frac{3\alpha}{4\pi}(4\pi)^{\epsilon}\,\Gamma(\epsilon)\;.
\eeq
Note that $\delta Z_2$ is gauge dependent, whereas $\delta Z_m$ is gauge independent. Plugging the counterterms into (\ref{eq:Ageneral}) and (\ref{eq:Bgeneral}),
\begin{align}
A(p^2)^{\overline{\rm MS}} &=\frac{\alpha\,\ahat}{4\pi}\left\{\left(1+\frac{m^2}{p^2}\right)\left[1-\left(1-\frac{m^2}{p^2}\right)\ln\left(1-\frac{p^2}{m^2}\right)\right]-\ln\left(\frac{m^2}{\mu^2}\right)\right\},\nonumber\\
B(p^2)^{\overline{\rm MS}} &= \frac{\alpha}{2\pi}\left\{2+\ahat-{1\over 2}(3+\ahat)\left[\left(1-\frac{m^2}{p^2}\right)\ln\left(1-\frac{p^2}{m^2}\right)+\ln\left(\frac{m^2}{\mu^2}\right)\right]\right\}\;.
\end{align}
The physical pole mass, denoted by  $m_e$, corresponds to a zero of the inverse propagator,
\begin{align}
\Gamma^{(2)}\of{p^2}\Big|_{\slashed{p} = m_e} = 0.
\end{align}
At one loop order, $\Gamma^{(2)}\of{p^2}$ is proportional to $\slashed{p}-m\bigl(1+B(p^2)^{\overline{\rm MS}} -A(p^2)^{\overline{\rm MS}} \bigr)$. Off-shell, the quantity $B(p^2)-A(p^2)$ depends on the gauge parameter,
\begin{align}
B(p^2)^{\overline{\rm MS}} -A(p^2)^{\overline{\rm MS}} =\frac{\alpha}{4\pi}\left\{4+\ahat\left(1-\frac{m^2}{p^2}\right)-3\ln\left(\frac{m^2-p^2}{\mu^2}\right)+\frac{m^2}{p^2}\left[3+\ahat\left(1-\frac{m^2}{p^2}\right)\right]\ln\left(1-\frac{p^2}{m^2}\right)\right\}.
\end{align}
The electron pole mass, however, depends on $B-A$ on-shell,
\begin{align}
m_e=m\bigl[1+B(m_e^2)^{\overline{\rm MS}} -A(m_e^2)^{\overline{\rm MS}} \bigr]\,.
\end{align}
One can easily check that $B(m^2)-A(m^2)$ is independent of $\ahat$, demonstrating the gauge invariance of the pole mass through one loop order.  Indeed, the pole mass must be IR finite and independent of the gauge parameter $\ahat$ to all orders in perturbation theory~\cite{Tarrach:1980up}.

\subsection{On-shell (OS) Renormalization}
It is also instructive to use the OS subtraction scheme,
 where the parameter $m$ is identified as the pole mass.  Here a well-known IR divergence appears in the electron wave function counterterm. This divergence is unrelated to the IR divergences in supersymmetric QED analyzed earlier in this paper, appearing only as an artifact of the OS renormalization scheme, but it is interesting to see how it -- and gauge dependence -- appear in the self-energy.  Writing 
\beq
\Sigma(p)=\Sigma(m)+(\slashed{p}-m)\Sigma^\prime(m)+\mathcal{O}\bigl((\slashed{p}-m)^2\bigr)\,,
\eeq
the OS renormalization conditions  are:
\beq \label{bcs}
\Sigma(m)^{\rm OS}=0\,,\qquad\quad \Sigma^\prime(m)^{\rm OS}=0\,.
\eeq
It then follows that the inverse propagator can be written as
\beqa
\Gamma^{(2)}(p)^{\rm OS}=\slashed{p}-m-\Sigma(p)^{\rm OS}&=&\bigl[1+\Sigma^\prime(m)^{\rm OS}\bigr](\slashed{p}-m)-\Sigma(m)^{\rm OS}+\mathcal{O}\bigl((\slashed{p}-m)^2\bigr)\nonumber \\[6pt]
&=&\slashed{p}-m+\mathcal{O}\bigl((\slashed{p}-m)^2\bigr)\,.\nonumber
\eeqa
Employing \eq{AB}, we can rewrite the boundary conditions specified in \eq{bcs} as
\beq \label{ABbc}
A(m^2)^{\rm OS}=B(m^2)^{\rm OS}\,,\qquad\quad 
A(m^2)^{\rm OS}=2m^2\left[\left(\frac{\partial B^{\rm OS}}{\partial p^2}\right)-\left(\frac{\partial A^{\rm OS}}{\partial p^2}\right)\right]_{p^2=m^2}\,,
\eeq
Using the first boundary condition and \eqs{eq:Ageneral}{eq:Bgeneral}, 
 we conclude that
\beq \label{zmos}
\delta Z_m^{\rm OS}=-\frac{\alpha}{4\pi}(4\pi)^{\epsilon}\Gamma(\epsilon)\left(\frac{m^2}{\mu^2}\right)^{-\epsilon}
 \left(\frac{3-2\epsilon}{1-2\epsilon}\right)
 =-\frac{3\alpha}{4\pi}\left[(4\pi)^\epsilon\Gamma(\epsilon)+\tfrac{4}{3}-\ln\left(\frac{m^2}{\mu^2}\right)\right]\,,\nonumber
 \eeq
 after dropping terms of $\mathcal{O}(\epsilon)$. Similarly, $\delta Z_2^{\rm OS}$ may be obtained from the second boundary condition,
 \beq \label{deltaz2}
\delta Z_2^{\rm OS}=-\frac{\alpha\,\ahat}{4\pi}\left(\frac{m^2}{\mu^2}\right)^{-\epsilon}\frac{(4\pi)^\epsilon \Gamma(\epsilon)}{1-2\epsilon}+\frac{\alpha}{4\pi}\left(\frac{m^2}{\mu^2}\right)^{-\epsilon}\frac{(4\pi)^\epsilon \Gamma(1+\epsilon)}{\epsilon(1-2\epsilon)}\bigl[\ahat-3+2\epsilon\bigr]\,.
\eeq
The term on the right hand side of \eq{deltaz2} proportional to $\Gamma(\epsilon)$ reflects the ultraviolet divergence in the unregulated self-energy integral [cf.~\eq{zees}].  The last term on the right hand side of \eq{deltaz2} which contains a pole at $\epsilon=0$ reflects a new infrared divergence, an artifact of the OS scheme choice.  (Note that this one-loop infrared divergence is absent in the Yennie gauge~\cite{Fried:1958zz,Tomozawa:1979xn,Karshenboim:1988nf,Adkins:1993qm,Adkins:1994uy,Eides:2001dw} , $\ahat=3$.)

\noindent We can determine $A(p^2)$ and $B(p^2)$ in the on-shell scheme by writing
\begin{align}
A(p^2)_{\rm OS}&=A(p^2)_{\overline{\rm MS}}+\delta Z_2^{\rm OS}-\delta Z_2^{\overline{\rm MS}}\\
B(p^2)_{\rm OS}&=B(p^2)_{\overline{\rm MS}}+\delta Z_m^{\rm OS}+\delta Z_2^{\rm OS}-\delta Z_m^{\overline{\rm MS}}-\delta Z_2^{\overline{\rm MS}}\,.
\end{align}
\Eqss{zees}{zmos}{deltaz2} yield,
\begin{align}
\delta Z_2^{\rm OS}-\delta Z_2^{\overline{\rm MS}}&=-\frac{\alpha\,\ahat}{2\pi}
\left[1-\tfrac12\ln\left(\frac{m^2}{\mu^2}\right)\right]+\frac{\alpha}{4\pi}\left(\frac{m^2}{\mu^2}\right)^{-\epsilon}\frac{(4\pi)^\epsilon \Gamma(1+\epsilon)}{\epsilon(1-2\epsilon)}\bigl[\ahat-3+2\epsilon\bigr]\label{diff2} \\
\delta Z_m^{\rm OS}-\delta Z_m^{\overline{\rm MS}}&=-\frac{\alpha}{\pi}\left[1-\tfrac34\ln\left(\frac{m^2}{\mu^2}\right)\right]\;.\label{diffm}
\end{align}
The infrared divergence is explicitly exhibited in \eq{diff2}.  Expanding about $\epsilon=0$ yields
\beq \label{z2diff}
\delta Z_2^{\rm OS}-\delta Z_2^{\overline{\rm MS}}=\frac{\alpha(\ahat-3)}{4\pi}(4\pi)^\epsilon \Gamma(\epsilon)-\frac{\alpha}{\pi}\left[1-\tfrac34\ln\left(\frac{m^2}{\mu^2}\right)\right]\,.
\eeq
Thus, both $A(p^2)_{\rm OS}$ and $B(p^2)_{\rm OS}$ are infrared divergent if $\ahat\neq 3$.
Note that the difference $B(p^2)_{\rm OS}-A(p^2)_{\rm OS}$ is infrared finite. As in the ${\overline{\rm MS}}$ scheme, $B(p^2)_{\rm OS}-A(p^2)_{\rm OS}$ depends on the gauge parameter $\ahat$ for general $p^2$, but becomes gauge invariant on-shell (vanishing by construction).

\section{The two-point functions of SQED}

\subsection{Tree-level propagators of SQED in a covariant gauge}

In this subsection, we focus on the terms of the SUSY-QED Lagrangian that are independent of the chiral superfields $\Phi_{\pm}$.   These terms are given by the last two terms of \eq{eq:SUSYQEDLagrangian}, which can also be written in the following
form,
\beq
\mathscr{L}_{\rm SQED}=\tfrac{1}{4}\bigl[\mathcal{W}^\alpha \mathcal{W}_\alpha\bigr]_{\theta\theta}+\tfrac{1}{4}\bigl[\overline{\mathcal{W}}_{\dalpha} \overline{\mathcal{W}}^{\dalpha}\bigr]_{\thetabar\mspace{1.5mu}\thetabar}-\tfrac{1}{8}\xi\bigl[(D^2 V)(\overline{D}^2 V)\bigr]_{\theta\theta\thetabar\mspace{1.5mu}\thetabar}\,,
\eeq
where the subscript $\theta\theta$ instructs one to take the coefficient of $\theta\theta$ of the corresponding superfield, etc., and the spinor chiral superfield $\mathcal{W}_\alpha$ is defined by
\beqa
\mathcal{W}_\alpha(x,\theta,\overline{\theta})&=&-\tfrac{1}{4}\overline{D}^2 D_\alpha V(x,\theta,\overline{\theta})  \\
&=&
\exp(-i\theta\sigma^\mu\overline{\theta}\partial_\mu)\biggl\{-i\lambda_\alpha(x)+\theta_\alpha D(x)-\half i(\sigma^\mu\sigmabar^\nu\theta)_\alpha F_{\mu\nu}(x)-\theta\theta[\sigma^\mu\partial_\mu\bar{\lambda}(x)]_\alpha\biggr\}\,,\nonumber
\eeqa
where $F_{\mu\nu}\equiv\partial_\mu V_\nu-\partial_\nu V_\mu$.

The two-component spinor notation employed in this paper follows that of Refs.~\cite{Sohnius:1985qm,Dreiner:2008tw}.
Following Ref.~\cite{Sohnius:1985qm}, the spinor covariant derivatives are given by
\beqa
D_\alpha&=&\frac{\partial}{\partial\theta^\alpha}-i\sigma^\mu_{\alpha\dbeta}\thetabar^{\dbeta}\partial_\mu\,, \\[6pt]
\overline{D}_{\dalpha}&=&-\frac{\partial}{\partial\thetabar^{\dalpha}}+i\theta^\beta \sigma^\mu_{\beta\dalpha}\partial_\mu\,.
\eeqa
Using these definitions,
\beqa
D^2\equiv D^\alpha D_\alpha&=& \epsilon^{\alpha\beta}D_\beta D_\alpha= -\partial^\alpha\partial_\alpha +2i\thetabar_{\dalpha}\sigmabar^{\mu\dalpha\beta}\partial_\beta \partial_\mu+\thetabar\mspace{1.5mu}\thetabar\,\Box\,,\\[6pt]
\overline{D}^2\equiv \overline{D}_{\dalpha} \overline{D}^{\dalpha}&=& 
\epsilon^{\dalpha\dbeta} \overline{D}_{\dalpha}\overline{D}_{\dbeta}=
-\partial_{\dalpha}\partial^{\dalpha} +2i\theta^\alpha\sigma^\mu_{\alpha\dbeta}\overline{\partial}^{\dbeta}\partial_\mu+\theta\theta\,\Box\,,
\eeqa
where $\Box\equiv \partial_\mu\partial^\mu$.

Hence, the super-QED Lagrangian including the gauge fixing term (after dropping a total derivative) is\footnote{The explicit form of the bosonic part of $\mathscr{L}_{\rm SQED}$ in the $R_\xi$ gauge can be found in Ref.~\cite{Miller:1983pg}.}
\beqa
\mathscr{L}_{\rm SQED}&=&-\tfrac{1}{4}F^{\mu\nu}F_{\mu\nu}-\tfrac{1}{2}\xi(\partial_\mu V^\mu)^2+\tfrac12\left(1-\xi\right)D^2
-\tfrac{1}{2}\xi\bigl[(\Box a)^2-2D\Box a+(\partial_\mu M)^2+(\partial_\mu N)^2\bigr]
\nonumber \\[6pt]
&& \qquad\qquad\qquad
 + i\left(1-\xi\right)\overline{\lambda}\sigmabar^\mu\partial_\mu\lambda-\xi\bigl[i\partial_\mu\overline{\chi}\,\sigmabar^\mu\Box\chi+\lambda\Box\chi+\overline{\lambda}\Box\overline{\chi}\bigr]\,.\label{lagsqed}
\eeqa

To compute the tree-level propagators $\vev{aa}$, $\vev{aD}$ and $\vev{DD}$, we write
\beq
\mathscr{L}_{\rm SQED}\ni \tfrac12\left(1-\xi\right)D^2
-\tfrac{1}{2}\xi\bigl[(\Box a)^2-2D\Box a\bigr]=\tfrac{1}{2}(a\quad D)\begin{pmatrix} -\xi\Box^2 & \quad \xi\Box \\ \xi\Box & \quad
1-\xi\end{pmatrix}\begin{pmatrix} a \\ D\end{pmatrix}\,.
\eeq
We compute the inverse,
\beq
\begin{pmatrix} -\xi\Box^2 & \quad \xi\Box\\ \xi\Box & \quad
1-\xi\end{pmatrix}^{-1}=\begin{pmatrix} (1-\xi^{-1})\Box^{-2} & \quad \Box^{-1} \\ \Box^{-1} & \quad
1\end{pmatrix}\,.
\eeq
We can also work in momentum space by taking $\partial_\mu\longrightarrow -ik_\mu$.  It then follows that the momentum space propagator matrix is,
\beq
i\Delta(k)=i\begin{pmatrix} (1-\xi^{-1})/k^4 & \quad  -1/k^2 \\  -1/k^2 & \quad 1\end{pmatrix}\,.
\eeq
Hence, 
\beq \label{prop0}
\vev{aa}=i(1-\xi^{-1})/k^4\,,\qquad\quad \vev{aD}=-i/k^2\,,\qquad\quad  \vev{DD}=i\,.
\eeq

The tree-level fermionic propagators are obtained by writing
\beqa \label{fmatrix}
&&
\mathscr{L}_{\rm SQED}\ni \half\bigl(\lambda\quad \overline{\lambda}\quad\chi\quad\overline{\chi}\bigr)\begin{pmatrix}
0 & i(1-\xi)\sigma^\mu\partial_\mu & \quad -\xi\Box & \quad 0 \\
i(1-\xi)\sigmabar^\mu\partial_\mu & \quad 0 & \quad 0 & -\xi\Box \\
-\xi\Box & \quad 0 & \quad 0 & \quad i\xi\sigma^\mu \partial_\mu \Box \\
0 & \quad -\xi & \quad i\xi\sigmabar^\mu\partial_\mu\Box & \quad 0\end{pmatrix}
\begin{pmatrix} \lambda \\ \overline{\lambda} \\ \chi \\ \overline{\chi}\end{pmatrix}\,,\nonumber \\
&&\phantom{line}
\eeqa
which differs from the fermionic part of \eq{lagsqed} by a total derivative which is subsequently dropped.
The inverse of the matrix that appears in \eq{fmatrix} is
\beq
-\,\frac{1}{\Box^2}\begin{pmatrix} 0 & \quad i\sigma^\mu\partial_\mu\Box & \quad \Box & \quad 0 \\
 i\sigmabar^\mu\partial_\mu\Box  & \quad 0 & \quad 0 & \quad \Box\\
 \Box & \quad 0 & \quad 0 & \quad i(\xi^{-1}-1)\sigma^\mu\partial_\mu \\
 0 & \quad \xi^{-1}\Box & \quad  i(\xi^{-1}-1)\sigmabar^\mu\partial_\mu & \quad 0 \end{pmatrix}\,.
 \eeq
 We can now read off the propagator matrix in momentum space by taking $\partial_\mu\longrightarrow -ik_\mu$,
 \beq
 i\Delta(k)=\frac{i}{k^2}\begin{pmatrix} 0 & \quad \sigma\newcdot k & \quad 1 & \quad 0 \\
 \sigmabar\newcdot k & \quad 0 & \quad 0 & 1 \\
 1 & \quad 0 & \quad 0 & \quad (1-\xi^{-1})\sigma\newcdot k/k^2 \\
 0 & \quad 1 & \quad (1-\xi^{-1})\sigmabar\newcdot k/k^2 & \quad 0\end{pmatrix}\,.
 \eeq
 It follows that
 \beqa
 &&
 \vev{\lambda\overline{\lambda}}=\frac{i\sigma\newcdot k}{k^2}\,,\qquad\qquad\qquad\quad 
 \vev{\overline{\lambda}\lambda}=\frac{i\sigmabar\newcdot k}{k^2}\, \label{prop1lambda} \\
 &&\vev{\chi\overline{\chi}}=\frac{i(1-\xi^{-1})\sigma\newcdot k}{k^4}\,,\qquad\quad \,
 \vev{\overline{\chi}\chi}=\frac{i(1-\xi^{-1})\sigmabar\newcdot k}{k^4}\, \label{prop2chi} \\
 &&\vev{\lambda\chi}=\vev{\chi\lambda}=\vev{\overline{\lambda}\overline{\chi}}\vev{\overline{\chi}\overline{\lambda}}=\frac{i}{k^2}\,,
 \label{prop3} \\
 &&\vev{\lambda\lambda}=\vev{\overline{\lambda}\,\overline{\lambda}}=\vev{\chi\chi}=\vev{\overline{\chi}\,\overline{\chi}}
 =\vev{\lambda\overline{\chi}}=\vev{\chi\overline{\lambda}}=\vev{\overline{\lambda}\chi}=\vev{\overline{\chi}\lambda}=0\,.
 \label{prop4}
 \eeqa
 Note that the propagators for the gauginos ($\lambda$ and $\overline{\lambda}$) are standard fermionic propagators for massless two-component fermions~\cite{Dreiner:2008tw}.  
 
Finally, the inverse of the terms quadratic in the vector boson fields is the well-known QED expression,
\beq
\frac{1}{\Box}\left(g_{\mu\nu}-(1-\xi^{-1})\frac{\partial_\mu\partial_\nu}{\Box}\right)\,.
\eeq
That is, in momentum space, we obtain the standard tree-level photon propagator in a covariant gauge,
\beq
\vev{V_\mu V_\nu}=\frac{i}{k^2}\left(-g_{\mu\nu}+(1-\xi^{-1})\frac{k_\mu k_\nu}{k^2}\right)\,.
\eeq
 
The tree-level propagators can be obtained directly from a single master formula written in terms of the vector superfield,\footnote{In this notation for the propagator, the time ordered product symbol $T$ is suppressed.}
\beqa 
&&\hspace{-0.15in}
\vev{V(x,\theta,\overline{\theta})V(y,\zeta,\overline{\zeta})}= \frac{i}{\Box}\exp\bigl[i(\theta\sigma^\mu\overline{\zeta}-\zeta\sigma^\mu\overline{\theta})\partial_\mu\bigr]\biggl\{\frac{1-\xi^{-1}}{\Box} 
+\tfrac{1}{4}(1+\xi^{-1})\delta^4(\theta-\zeta)\biggr\}\delta^4(x-y)\,,\nn \\
 &&\phantom{line}\label{master}
\eeqa
where
\beq
 \delta^4(\theta-\zeta)\equiv
 (\theta-\zeta)^\alpha(\theta-\zeta)_\alpha\,(\overline{\theta}-\overline{\zeta})_{\dbeta}(\overline{\theta}-\overline{\zeta})^{\dbeta}\,.
 \eeq
 In momentum space, \eq{master} yields,
 \beq
 \vev{V(\theta,\overline\theta)V(\zeta,\overline\zeta)}=\frac{i}{k^2}\exp\bigl[\theta \sigma\newcdot k\,\overline{\zeta}-\zeta\sigma\newcdot k\,\overline \theta\,\bigr]\biggl\{\frac{1-\xi^{-1}}{k^2}-\tfrac{1}{4}(1+\xi^{-1})\delta^4(\theta-\zeta)\biggr\},
 \eeq
which is the result quoted in \eq{vpropagator}.  It is straightforward to check that \eq{master} reproduces the
tree-level propagators of the component fields obtained above.

The renormalization of SQED coupled to matter is highly non-trivial, in light of the fact that the supersymmetric gauge-invariant Lagrangian is inherently non-linear.  Supersymmetric procedures for the renormalization of gauge theories (that do not impose the Wess-Zumino gauge) have been proposed in Ref.~\cite{Slavnov:1974uv}.

\subsection{Relations among the SQED two-point functions}

Consider a chiral supermultiplet, 
\beq
\Phi(x,\theta,\thetabar) = \exp(-i\theta\sigma^\mu\thetabar\partial_\mu)\bigl[
\phi(x) + \sqrt{2}\,\theta \psi(x) + \theta\theta F(x)\bigr]\,.
\eeq
The component fields transform as
\beq
\delta_\eta \phi &=& \sqrt{2}\,\eta \psi\,,\nn\\
\delta_\eta\psi_\alpha
&=&
- i\sqrt{2} (\sigma^\mu \overline\eta)_\alpha\> \partial_\mu \phi+\sqrt{2}\,\eta_\alpha F\,,\nn \\
\delta_\eta F&=&-i\sqrt{2}\,\overline\eta\,\sigmabar^\mu\partial_\mu\psi\,,
\eeq
where $\eta$ and $\overline{\eta}$ are anticommuting parameters.
By hermitian conjugation, 
\beqa
\delta_\eta \phi^* &= &\sqrt{2}\,\overline\eta \overline \psi\,,\nn \\
\delta_\eta\overline{\psi}_{\dot{\alpha}}
&=&
 i \sqrt{2}(\eta\sigma^\mu)_{\dot{\alpha}}\>   \partial_\mu \phi^*+\sqrt{2}\,\overline\eta_{\dot\alpha}F^*\,,\nn \\
\delta_\eta F^*&=&i\sqrt{2}(\partial_\mu\overline\psi)\sigmabar^\mu\eta\,.
 \eeqa
 The transformed fields $A=\phi$, $\psi$ or $F$  (or their corresponding complex conjugated fields) can be expressed in terms of the commutators
 \beq \label{dxi}
 \deltaxi A(x)=i\bigl[\eta Q+\overline{\eta}\overline{Q}\,,\,A(x)\bigr]\,,
 \eeq
 where $Q$ and $\overline{Q}$ generate supersymmetric translations.  
 
Consider first the identity
\beq
i\langle 0|\bigl[\eta Q+\overline{\eta}\overline{Q}\,,\,\psi_\alpha(x)\phi(y)\bigr]|0\rangle=0\,,
\eeq
which follows under the assumption that the supersymmetry generators annihilate the vacuum (i.e., supersymmetry is an unbroken symmetry).   In light of \eq{dxi}, we obtain,
\beqa
0&=&\langle 0|\deltaxi\bigl[\psi_\alpha(x)\phi(y)\bigr]|0\rangle=
\langle 0|\bigl[\psi_\alpha(x)+\deltaxi\psi_\alpha(x)\bigr]\bigl[\phi(y)+\deltaxi\phi(y)\bigr]-\psi_\alpha(x)\phi(y)|0\rangle
\nn \\
&=&\langle 0|[\deltaxi\psi_\alpha(x)]\phi(y)+\psi_\alpha(x)[\deltaxi\phi(y)]|0\rangle\,.
\eeqa
Plugging in the transformation laws given above,
\beq \label{eq1}
\eta^\beta\biggl\{ \epsilon_{\alpha\beta} \langle 0| F(x)\phi(y)|0\rangle-\langle 0|\psi_\alpha(x)\psi_\beta(y)|0\rangle\biggr\}
-i\overline\eta^{\dot\beta}\sigma^\mu_{\alpha\dot{\beta}}\partial^x_\mu \langle 0|\phi(x)\,\phi(y)|0\rangle =0\,,
\eeq
where $\partial^x_\mu\equiv \partial/\partial x^\mu$.  
The coefficients of $\eta$ and $\overline{\eta}$ must separately vanish.  Thus, we conclude that\footnote{Note that \eq{eq1} also implies that
 $\langle 0|\phi(x)\,\phi(y)|0\rangle$ is a constant (independent of position) after noting that the 2-point function is translationally invariant.}
\beq \label{id1}
\langle 0|\psi_\alpha(x)\psi_\beta(y)|0\rangle=\epsilon_{\alpha\beta} \langle 0| F(x)\phi(y)|0\rangle\,.
\eeq

Similarly, the identity,
\beq
i\langle 0|\bigl[\eta Q+\overline{\eta}\overline{Q}\,,\phi^*(x)\overline\psi_{\dot\beta}(x)\bigr]|0\rangle=0\,.
\eeq
yields
\beq \label{idalt}
\langle 0|\overline\psi_{\dot\alpha}(x)\overline\psi_{\dot\beta}(y)|0\rangle=-\epsilon_{\dot\alpha\dot\beta} \langle 0| \phi^*(x)F^*(y)|0\rangle\,.
\eeq

Next, we consider the identity, 
\beq
i\langle 0|\bigl[\eta Q+\overline{\eta}\overline{Q}\,,\psi_\alpha(x)\phi^*(y)\bigr]|0\rangle=0\,.
\eeq
A similar computation yields,
\beq \label{eq2}
\overline{\eta}^{\dot\beta}\biggl\{-i\sigma^\mu_{\alpha\dot\beta}\partial_\mu \langle 0|\phi(x)\phi^*(y)|0\rangle+
\langle 0| \psi_\alpha(x)\overline{\psi}_{\dot\beta}(y) |0\rangle\biggr\}+\eta_\alpha \langle 0| F(x)\phi^*(y) |0\rangle\,.
\eeq
It follows that\footnote{Note that \eq{eq2} also implies that
 $\langle 0|F(x)\,\phi^*(y)|0\rangle=0$.}
\beq \label{eq3}
\langle 0| \psi_\alpha(x)\overline{\psi}_{\dot\beta}(y) |0\rangle=i\sigma^\mu_{\alpha\dot\beta}\partial_\mu  \langle 0|\phi(x)\phi^*(y)|0\rangle\,.
\eeq
It is convenient rewrite \eq{eq3} in momentum space,\footnote{\label{footnotefnft}%
The Fourier transform of a translationally invariant
function $f(x,y)\equiv f(x-y)$ is given by
$$
f(x,y)=\int\frac{d^4 p}{(2\pi)^4}\,\widehat f(p)\,e^{- ip\newcdot (x-y)}
\,,\qquad {\rm where} \qquad
\widehat f(p) = \int d^4 x\, f(x,0) e^{ ip\newcdot x}\,.
$$
In the notation of the text above,
$f(x,y)\ls{\rm FT}\equiv\widehat f(p)$. Moreover, we note that $f(y,x)_{\rm FT}=\widehat{f}(-p)$.\label{fn}}  
\beq \label{eq3II}
\langle 0| \psi_\alpha(x)\overline{\psi}_{\dot\beta}(y) |0\rangle_{\rm FT}=p\newcdot\sigma_{\alpha\dot\beta}\langle 0|\phi(x)\phi^*(y)|0\rangle\,.
\eeq

Finally, we consider the identity, 
\beq
i\langle 0|\bigl[\eta Q+\overline{\eta}\overline{Q}\,,\,F^*(x)\psi_\beta(y)\bigr]|0\rangle=0\,.
\eeq
Once again, a similar computation yields,
\beq \label{eq4}
\!\!\!\!\!\!
\biggl\{\delta^\alpha_\beta\langle 0|F^*(x) F(y)|0\rangle+i\sigmabar^{\mu\dot{\alpha}\alpha}\langle 0|(\partial_\mu\overline\psi_{\dot\alpha}(x)\psi_\beta(y)|0\rangle\biggr\}\eta_\alpha-i\overline\eta^{\dot\alpha}\sigma^\mu_{\beta\dot\alpha}\partial^y_\mu\langle 0|F^*(x)\phi(y)|0\rangle=0.
\eeq
It follows that
\beq \label{eq5}
\delta_\beta^\alpha \langle 0|F^*(x) F(y)|0\rangle =-i\sigmabar^{\mu\dot\alpha \alpha}\partial_\mu\langle 0|\overline\psi_{\dot\alpha}(x)\psi_\beta(y)|0\rangle\,.
\eeq
After raising the spinor indices, we can manipulate \eq{eq5} into the following form,
\beq \label{id2}
\langle 0|(\overline\psi^{\dot\alpha}(x)\psi^\beta(y)|0\rangle=\frac{i\sigmabar^{\mu\dot\alpha\beta}\partial_\mu}{\square} \langle 0|F^*(x) F(y)|0\rangle\,.
\eeq
In momentum space, \eq{id2} takes the following form,
\beq \label{id3}
\langle 0|\overline\psi^{\dot\alpha}(x)\psi^\beta(y)|0\rangle_{\rm FT}=\frac{p\newcdot\sigmabar^{\dot\alpha\beta}}{p^2} \langle 0|F^*(x) F(y)|0\rangle_{\rm FT}\,.
\eeq

One further relation of interest can be found by comparing \eqs{eq3}{id3}.  In particular,
if we lower the spinor indices in \eq{id3}, anticommute the two fermion fields and interchange the position coordinates, then it follows that
\beq
\langle 0|\psi_\beta(x)\overline\psi_{\dot\alpha}(y))|0\rangle_{\rm FT}=\frac{p\newcdot\sigma_{\beta\dot\alpha}}{p^2} \langle 0|F^*(x) F(y)|0\rangle_{\rm FT}\,.
\eeq
Hence,
\beq \label{id4}
 \langle 0|F^*(x) F(y)|0\rangle_{\rm FT}=p^2\langle 0|\phi(x) \phi^*(y)|0\rangle_{\rm FT}\,.
 \eeq

Note that the supersymmetric relations obtained above also apply to the corresponding time-ordered 2-point functions; i.e., they apply to the corresponding propagators to all orders in perturbation theory.\footnote{Strictly speaking, we should make use of the $T^*$-product which has the property that one can freely move total derivatives from inside of the vacuum expectation value of the product of fields to outside.  This is equivalent to defining the $T$-product via its functional integral representation.}   
More general supersymmetric Ward identities that relate two and three-point 1PI Green functions in SQED can be found in Ref.~\cite{Walker:1999bp}.

To make contact with the analysis of Appendix A, we convert from two-component to four-component fermion notation.\footnote{The relation between the two-component spinor and four-component spinor notation is discussed in Ref.~\cite{Dreiner:2008tw}.}
  The four-component propagator function for fermions in momentum space is given by 
\beq 
\!\!\!\!\!\!
 \langle 0| T\Psi(x)\overline\Psi(y)|0\rangle_{\rm FT} =\
\begin{pmatrix} \langle 0| T \psi_\alpha(x)\psi^\beta(y)|0\rangle_{\rm FT} & \quad
 \langle 0| T \psi_\alpha(x)\overline\psi_{\dot\beta}(y)|0\rangle_{\rm FT}  \\[15pt]
 \langle 0| T \overline\psi^{\dot\alpha}(x)\psi^\beta(y)|0\rangle_{\rm FT} & \quad
 \langle 0| T \overline\psi^{\dot\alpha}(x)\overline\psi_{\dot\beta}(y)|0\rangle_{\rm FT} \end{pmatrix},
 \eeq
where $\Psi(x)$ is a four-component spinor.  Using the supersymmetric relations obtained above, it
 follows that
 \beq   \label{prop1}
 \!\!\!\!\!\!
 \langle 0| T\Psi(x)\overline\Psi(y)|0\rangle_{\rm FT} =\begin{pmatrix} -\delta_\alpha{}^\beta\langle 0| T F(x)\phi(y)|0\rangle_{\rm FT} & \quad \displaystyle\frac{p\newcdot\sigma_{\alpha\dot\beta}}{p^2}\langle 0| T F^*(x)F(y)|0\rangle_{\rm FT}  \\[15pt]
 p\newcdot\sigmabar^{\dot\alpha\beta}\langle 0| T \phi(x)\phi^*(y)|0\rangle_{\rm FT}  & \quad
-\delta^{\dot\alpha}{}_{\dot\beta}\langle 0| T \phi^*(x)F^*(y)|0\rangle_{\rm FT}\end{pmatrix}.
\eeq

As a check, we apply the above results to SQED.   The matter fields correspond to two chiral multiplets, $\Phi_+=(\phi_+\,,\,\psi_+\,,\,F_+)$ and $\Phi_-=(\phi_-\,,\,\psi_-\,,\,F_-)$ with the corresponding superpotential, $W=m\Phi_+\Phi_-$.  The scalar field contributions to the Lagrangian are:
\beqa
\mathscr{L}&=&|\partial_\mu\phi_+|^2+|\partial_\mu\phi_-|^2+|F_+|^2+|F_-|^2+m(F_+\phi_-+F_-\phi_++{\rm h.c.})\nonumber \\[6pt]
 &=& \ofp{ \begin{array}{cc} F_+^*  & \quad\phi_-  \end{array} } 
\Delta_{0F\phi}^{-1}
\ofp{ \begin{array}{c}  F_+ \\[6pt] \phi_-^*  \end{array} }  + \ofp{ +\longleftrightarrow -  }\,,
\eeqa
where we have dropped terms that are a total derivative and we have defined the inverse tree-level propagator matrix
\begin{align} \label{dfi}
\Delta_{0F\phi}^{-1} \equiv  \ofp{ \begin{array}{cc} 1  & \quad m \\
m & \quad -\square \end{array} }.
\end{align}
Inverting this matrix and passing to momentum space yields
\beq \label{df}
i\Delta_{0F\phi}(p)=\frac{i}{p^2-m^2}\begin{pmatrix}
 \phm p^2 & \,\,\, -m \\  -m & \,\,\, \phm 1\end{pmatrix}
 \eeq
 
 Defining the Dirac electron field by
\beq
\Psi=\begin{pmatrix} \psi_+ \\[6pt] \overline{\psi}_-\end{pmatrix},
\eeq
the four-component electron propagator is given by
\beq  \label{prop2}
\!\!\!\!\!\!\!\!
 \langle 0| T\Psi(x)\overline\Psi(y)|0\rangle_{\rm FT} =\begin{pmatrix} -\delta_\alpha{}^\beta\langle 0| T F_+(x)\phi_-(y)|0\rangle_{\rm FT} & \quad \displaystyle\frac{p\newcdot\sigma_{\alpha\dot\beta}}{p^2}\langle 0| T F^*_+(x)F_+(y)|0\rangle_{\rm FT}  \\[15pt]
 p\newcdot\sigmabar^{\dot\alpha\beta}\langle 0| T \phi_-(x)\phi^*_-(y)|0\rangle_{\rm FT}  & \quad
-\delta^{\dot\alpha}{}_{\dot\beta}\langle 0| T \phi_-^*(x)F_+^*(y)|0\rangle_{\rm FT}\end{pmatrix}\!\!.
\eeq
In light of \eq{df}, we end up with the usual tree-level electron propagator,
\beq
\langle 0| T\Psi(x)\overline\Psi(y)|0\rangle_{\rm FT} ^{\rm tree}=\frac{i(\slashchar{p}+m)}{p^2-m^2+i\varepsilon}\,.
\eeq
In principle, the radiatively-corrected electron pole mass is obtained by inverting the $4\times 4$ propagator matrix given by \eq{prop2}, computing its determinant and finding the value of $p^2$ at which the determinant vanishes (details can be found in Ref.~\cite{Martin:2005ch}).  However, it is significantly simpler to perform the computations by analyzing the radiative corrections in the scalar ($\phi$--$F$) sector, as discussed in 
Sect.~\ref{perturbationtheorystructure}.

\section{Cancellation of Finite Corrections}
\label{finitecorrections}

In the text, we focused on cancellation of the leading IR divergent pieces from the correction to the physical electron mass in SQED with ${\overline{\rm DR}}$ renormalization. Here we demonstrate that the finite pieces also cancel.

Using dimensional regularization with $d= 4-2\delta$, $\delta <0$,
the correction to the mass term, given by (\ref{eq:1scalar_integral_form}), may be expressed
as
\begin{align}
I_{F_+ \phi_- } + \mathrm{c.t.} =   - 2 m \coupling^2   \ofp{ 1 - {1\over \xi}} {   p^2 \over p^2 - m^2 } {1\over \ofp{ 4 \pi}^2 } \ofp{ 4 \pi}^\epsIR  { \Gamma\of{\epsIR} \over 1 - \epsIR} \ F\of{  1+\epsIR,-\epsIR;2-\epsIR;{ p^2 \over p^2 - m^2 } }  { 1 \over \ofp{ m^2 - p^2 }^\epsIR } ,
\end{align}
where $F\equiv {}_2F_1$ is the Gauss hypergeometric function.
The correction to the kinetic term at leading order in $e$ is given by (\ref{eq:2scalar_integral_form}). The UV divergence gets canceled by the counterterm in ${\overline{\rm DR}}$, leaving only the finite and IR divergent pieces:
 \begin{align}
\begin{split}
I_{F_+^* F_+}^a + \mathrm{c.t.} 
= & { -2\coupling^2 \over \ofp{ 4 \pi}^2 } \ofp{ 1 - {1\over \xi}} 
  {  \ofp{ 4 \pi}^\epsIR \over \ofp{m^2 - p^2 }^\epsIR}  { p^2 \over p^2 - m^2 } { 2 \Gamma\of{ \epsIR} \over \ofp{ 2- \epsIR} \ofp{ 1-\epsIR}  } F\of{ 1 + \epsIR, -\epsIR;3-\epsIR;{ p^2 \over p^2 - m^2 }}.
\end{split}
\end{align}
The quantity relevant to the physical mass correction is therefore given by
\beqa
{I_{F_+ \phi_-} \over m} - I_{F_+^* F_+} &=& 
{ - 2 \coupling^2 \ofp{1-1/\xi} \over \ofp{ 4\pi}^2 }  { p^2 \over p^2 - m^2 } 
  { \ofp{ 4 \pi}^\epsIR \over \ofp{ m^2 - p^2 }^\epsIR } { \Gamma\of{\epsIR} \over 1-\epsIR} \\
&&\times 
   \cuof{ F\of{1+\epsIR,-\epsIR;2-\epsIR; {p^2 \over p^2 - m^2 } } - { 2 \over 2-\epsIR} F\of{1+\epsIR,-\epsIR; 3-\epsIR;{p^2\over p^2 - m^2 }} }.\nonumber
   \eeqa
The goal of this exercise is to evaluate the expression,
\beq
\!\!\!\!\!\!
\mathcal{F}(\epsilon)=\frac{\Gamma(\epsilon)}{1-\epsilon}\left\{F\left(1+\epsilon,-\epsilon\,;\,2-\epsilon,\frac{p^2}{p^2-m^2}\right)
-\frac{2}{2-\epsilon}F\left(1+\epsilon,-\epsilon\,;\,3-\epsilon,\frac{p^2}{p^2-m^2}\right)\right\}\!,
 \label{laurel}
\eeq
in the limit of $\epsilon\to 0$.

The relevant formulae taken from Ref.~\cite{bateman} are as follows.  First,  we make use of formula (42) on p.~103 of Ref.~\cite{bateman},
\beq
(c-b-1)F(a,b;c;z)+bF(a,b+1;c,z)-(c-1)F(a,b;c-1,z)=0\,.
\eeq
Choosing $a=1+\epsilon$, $b=-\epsilon$, $c=3-\epsilon$, it follows that
\beq
(2-\epsilon)F(1+\epsilon,-\epsilon\,;\,2-\epsilon,z)-2F(1+\epsilon,-\epsilon\,;\,3-\epsilon;z)=-\epsilon
F(1+\epsilon,1-\epsilon\,;\,3-\epsilon;z)\,.
\eeq
Using this result in \ref{laurel} with $z\equiv p^2/(p^2-m^2)$ yields,
\beq
\mathcal{F}(\epsilon)=-\frac{\Gamma(1+\epsilon)}{(1-\epsilon)(2-\epsilon)}F\left(1+\epsilon,1-\epsilon\,;\,3-\epsilon;\frac{p^2}{p^2-m^2}\right)\,,
\eeq
after using $\epsilon\Gamma(\epsilon)=\Gamma(1+\epsilon)$.  Taking the $\epsilon\to 0$ limit, we end up with
\beq
\mathcal{F}(0)=-\tfrac{1}{2}F\left(1,1;3;\frac{p^2}{p^2-m^2}\right)\,.
\eeq

Next, we make use of formula (15) on p.~102 of Ref.~\cite{bateman}, which implies that
\beq
F(1,1;2;z)=-\frac{\ln(1-z)}{z}\,.
\eeq
We then use formula (24) of p.~102  of Ref.~\cite{bateman} which gives (for $n=1$),
\beq
\frac{(c-a)(c-b)}{c}(1-z)^{a+b-c-1}F(a,b;c+1;z)=\frac{d}{dz}\biggl[(1-z)^{a+b-c}F(a,b;c;z)\biggr]\,,
\eeq
to derive
\beq \label{F}
F(1,1;3;z)=\frac{2}{z}\left[1+\frac{(1-z)\ln(1-z)}{z}\right]\,.
\eeq
Making use of \eq{F}, we arrive at our final result,
\beq
\mathcal{F}(0)=\left(\frac{m^2}{p^2}-1\right)\left[1+\frac{m^2}{p^2}\ln\left(1-\frac{p^2}{m^2}\right)\right]\,.
\eeq
The limit of $p^2\to m^2$ then yields,
\beq
\lim_{p^2\to m^2}\mathcal{F}(0)=0\,.
\eeq
Thus, finite corrections to the physical mass vanish in the on-shell limit.

\bibliographystyle{JHEP}

\end{document}